\RequirePackage{fix-cm}
\documentclass[twocolumn,epjc3]{svjour3}  
\smartqed  
\RequirePackage{graphicx}
\journalname{Eur. Phys. J. C}

\usepackage[dvipsnames]{xcolor}
\usepackage{amssymb}
\usepackage{cite}
\usepackage{amsmath}
\usepackage[normalem]{ulem}
\usepackage{soul}

\begin{document}

\title{On the breaking of Casimir scaling in jet quenching
}


\author{ Liliana Apolin\'{a}rio\thanksref{e1,addr1, addr2}
        \and
        Jo\~{a}o Barata \thanksref{e2,addr3,addr4}
        \and
        Guilherme Milhano\thanksref{e3,addr1, addr2} 
}

\thankstext{e1}{liliana@lip.pt}
\thankstext{e3}{joao.barata@cern.ch}
\thankstext{e2}{gmilhano@lip.pt}


\institute{LIP, Av. Prof. Gama Pinto, 2, P-1649-003 Lisboa, Portugal \label{addr1}
           \and
           Instituto Superior T\'{e}cnico (IST), Universidade de Lisboa, Av. Rovisco Pais 1, 1049-001, Lisboa,
Portugal \label{addr2}
           \and
           Instituto Galego de F\'{i}sica de Altas Enerx\'{i}as (IGFAE), Universidade de Santiago de Compostela,
E-15782 Galicia, Spain \label{addr3}
          \and
          Physics Department, Brookhaven National Laboratory, Upton, NY 11973, USA\label{addr4}
}

\date{Received: date / Accepted: date}

\maketitle

\begin{abstract}
We study the breaking of Casimir scaling, $C_F/C_A$, due to the evolution of jets in a hot and extended medium. By using JEWEL, a medium modified Monte Carlo event generator validated for a wide set of observables, we are able to study separately the development of quark and gluon-initiated jets produced back-to-back with a $Z-$boson. Focusing on the $p_T$ distribution, we conclude first that the colour charge ratio is constant but larger than Casimir scaling for high $p_T$ jets. In addition, our results seem to indicate that the medium induced cascade is more similar between quarks and gluons, thus driving the overall medium shower scaling away from the vacuum expectation. Finally, we compare our results to another Monte Carlo generator and observe the same scaling violation.
\end{abstract}

\section{Introduction}
\label{sec:intro}

Jet quenching, the modification of jet properties due to interactions with an underlying QCD medium, is one of the significant discoveries of the Heavy Ions Physics program at RHIC \cite{RHIC1, RHIC2} and the LHC \cite{LHC1, LHC2, LHC3}.\par 

The observed suppression of the yield of high $p_T$ jets points towards a picture where depleted energy is carried by soft radiation due to the interaction with a hot and extended medium \cite{Kurkela&Wiedeman,BlaizotMehtarFi}. For a single parton, such interactions correspond to small transverse momentum exchanges between the parton and the medium and a continuous colour rotation of the parton's colour charge. \par 

In the multiple soft scattering approximation (i.e. when the mean free path between two scattering centres is much smaller than the formation time of the radiated parton), the pattern of radiation emitted by a highly energetic in-medium parton is captured by the BDMPS-Z spectrum \cite{BDMPS1,BDMPS2,BDMPS3,BDMPS4,BDMPS5}. In the BDMPS-Z regime, the spectrum of emission is dominated by gluons of energy $\omega$ such that $\omega_{BH}\ll \omega < \omega_s \ll \omega_c < E$ with $E$ the energy of the initiating parton. Here $\omega_c\sim\hat{q}L^2$ is the frequency of gluons with a lifetime equal to the medium length $L$,  $\omega_s\sim\alpha_s\omega_c$ the frequency associated to the time after which most of the large angle gluons have been emitted \cite{BlaizotYacine} and $\omega_{BH}$ is the frequency cut before the single scattering regime becomes dominant.\par 

Although emissions with $\omega \sim \omega_c$ can have a significant impact on the average lost energy, it turns out that such energetic emissions are rare ($\mathcal{O}(\alpha_s)$). In fact, the spectrum is highly suppressed for emissions with $\omega > \omega_c$, a phenomenon that is known as the QCD Landau-Pomeranchuk-Migdal (LPM) effect\cite{LPM1,LPM2}. The main contribution comes from gluons with  $\omega \sim \omega_s$ that are emitted with a probability of $\mathcal{O}(1)$ \cite{BlaizotIancuDominguezYacine}. These gluons are independently emitted at any point in the medium shower and populate the large-angle region. \par

The LPM suppression, combined with the proliferation of soft wide-angle gluons, leads to a broadening of the jet structure, to the energy depletion of the jet core and, consequently, to a direct modification of the colour charge of the jet.\par 

Gluons emitted in the region $\omega_{BH}\ll \omega_s < \omega \ll \omega_c$ appear at smaller angles (see \cite{KonradYacine1}) than the previous soft gluons \cite{CMSangles}. Nonetheless, they correspond to rather rare emissions (O($\alpha_s$)), and therefore only contribute to higher-order in perturbation theory. Their contribution to the breaking of Casimir scaling should be, therefore, negligible.\par 

For the results shown in this paper, we use the fact that jets lose energy according to
\begin{equation}\label{eq:energyloss}
\left(\frac{dE_i}{dx}\right)^{BDMPS-Z}\propto C_i \, x  \, ,
\end{equation}
where $x$ is the longitudinal coordinate that the incoming particle has travelled. If the jet is fully coherent, $E_i$ and $C_i$ will refer to the energy and Casimir colour factor of the initiating parton. In the limit of fully decoherent jets, eq.\eqref{eq:energyloss} applies to each parton jet individually. The behaviour for a realistic jet, where some but not all partons are resolved by the medium, will be in between these two limits.\par 

In this manuscript, we investigate the role of the colour charge of the incoming parton in the development of its subsequent parton shower. For that purpose, we use two different Monte Carlo event generators that account for QCD jet evolution in heavy-ion collisions: one based on a perturbative prescription to calculate in-medium radiation (JEWEL\cite{Zapp:2011,Zapp:2012}) and another whose energy loss of partons is based on holographic calculations (Hybrid Strong/Weak coupling approach\cite{Hybrid3}). This allows to establish two benchmarks for the effect of colour charge in jet energy loss.

By using $Z +$ jet events, we can select quark and gluon-initiated jets, with the information of their initial transverse momentum scale (provided that the $Z-$boson transverse momentum is a good proxy for the back-to-back parton). We parallel the results obtained from reconstructed jets in vacuum and in the presence of an expanding medium as provided by the Monte Carlo model. We further compare the Monte Carlo results to the analytical expectations from eq.\eqref{eq:energyloss}. Our results are always presented in terms of a quark to gluon ratio. In vacuum, for the observables discussed and at lowest order, this ratio obeys the well known scaling $C_F/C_A$ (Casimir scaling). We study how this scaling gets modified in the presence of in-medium radiation.\par 

The paper is divided as follows. Section \ref{sec:1} presents the general set up of our work, section \ref{sec:2} shows the results obtained for the breaking of Casimir scaling in equation \eqref{eq:energyloss} and for the full in-medium jet. Finally section \ref{sec:4} gives a general discussion of our results.

\section{Set up} \label{sec:1}

To generate the samples needed for our study, we use the two Monte Carlo event generators mentioned in section \ref{sec:intro}. Both are able to reasonably describe most jet observables measured so far, in particular  those we consider in this work.\par

To access the initial $p_T$ scale of the produced jet and to have a handle on the jet flavour, we generate $Z+$quark-jet and $Z+$gluon-jet samples separately. This choice is well suited for our study since it allows a clear division in the flavour content of the samples and allows us to use the $p_T$ of the formed $Z-$boson as a proxy for the transverse momentum scale of the initiating parton.\par 

The majority of our study relies on ideal samples with no initial state radiation (ISR), no hadronization and no medium response. This way, we maximize the cleanest setup for colour charge studies, simplifying a comparison with analytical results. We afterwards check the effect of a more realistic situation, by considering hadronization and ISR effects separately. The collision is always considered to be at a centre-of-mass energy per nucleon pair of $\sqrt{s_{NN}}=5.02 $~TeV. As a reference for the full in-medium jet propagation, we take the $[0 - 10] \%$ centrality class, for the case of in-medium jets, although we also study the evolution with different centrality classes.\par 

We use JEWEL's default parameters ($\tau_i = 0.4$ fm, $T_i = 590$ MeV), known to capture jet and nuclear modification factor and Z+jet asymmetry at these energies \cite{KunnawalkamElayavalli:2016ttl,KunnawalkamElayavalli:2017hxo}, and generate 1 million weighted events for the four controlled samples: vacuum quark/gluon initiated jets, and in-medium quark/gluon initiated jets. For the centrality study, we produced $500$ thousand events for each additional centrality class: $[10-20] \%$ and $[30-50] \%$.\par

As for the hybrid Monte Carlo, the samples are generated at particle level with ISR. The model has one free parameter $\kappa=0.4$, which is fixed via fitting to data \cite{Hybrid3}. The simulated events corresponds to the $[0-10] \%$ centrality class. We generate roughly 3 million events (both for vaccum and medium), not separated in flavour. Therefore, there is a great imbalance between the quark and gluon samples. This only affects the magnitude of the statistical errors obtained.\par 
For the analysis cuts, we take the CMS $Z+$jet analysis procedure \cite{Sirunyan:2017jic}, adapting when necessary:
\begin{itemize}
  \item The reconstructed $Z-$boson, decaying into a muon pair, was required to have a transverse momentum $p_T^Z> 50$~GeV, a pseudo-rapidity $\mid \eta^Z\mid< 3$, and a mass within $70 < m_Z < 100$~GeV;
  \item Jets were reconstructed from all final state particles (after removing the muons from the particle list) with the anti-$k_t$ algorithm \cite{Antikt} for radius $R=0.2 - 0.5$, a transverse momentum of $p_T> 20 \ GeV$ and pseudo-rapidity $\mid \eta \mid<3$; 
  \item To find the back-to-back jet, we kept the hardest jet that is the azimuthal range from the $Z-$boson by $5 \pi /6<\mid \phi^Z -\phi^{jet} \mid<7 \pi/6$;
  \item The event is accepted if the jet satisfies $\mid\eta^{jet}\mid< 2.5$ (jet cone inside full $\eta$ acceptance region); otherwise, we reject the event. 
\end{itemize}
Jets reconstruction is done using FASTJET \cite{Fastjet}.\par 

In our analysis, for convenience, we introduce the following variable
\begin{equation}\label{eq:R}
R(\mathcal{O})_{med} =\frac{(\langle \mathcal{O}_ {vac}\rangle-\langle \mathcal{O}_{vac + med}\rangle)^{quark}}{(\langle \mathcal{O}_{vac}\rangle-\langle \mathcal{O}_{vac + med}\rangle)^{gluon}}
\end{equation}
where the lower indices $vac$ indicate the vacuum reference of each Monte Carlo model and $vac + med$ the full Monte Carlo simulation that naturally includes both vacuum-like and medium-induced radiation. The upper index specifies the sample flavour. The $\langle \ \rangle$ denotes the fact that we will use the average value of the observable in some bin (typically a $p_T$ bin). This allow us to isolate pure medium-induced emissions in the definition of $R(\mathcal{O})_{med}$.\par 

We further introduce $R(\mathcal{O})_i$ for the case where we only take each simulation ratios.
\begin{equation}
R(\mathcal{O})_i=\left(\frac{\langle \mathcal{O}\rangle^{quark}}{\langle \mathcal{O}\rangle^{gluon}}\right)_i \quad , \ i=vac,vac+med
\end{equation}

\section{Results}\label{sec:2}

In order to probe the color factor for jets we begin by exploring the $p_T$ distribution of the (identified) leading jet in the event.\par 
As argued in the previous section, the jet energy is depleted due to medium interactions and from equation (\ref{eq:energyloss}) one expects that for fully coherent jets the quark to gluon jet ratio would obey 
\begin{equation}\label{eq:energyratio}
\frac{dE_{quark}}{dx}\bigg/ \frac{dE_{gluon}}{dx}=\frac{C_F}{C_A}=\frac{4}{9}= 0.4444 \cdots    
\end{equation}
where $C_A=N$ and $C_F=(N^2-1)/2N$ for a general SU(N) group ($N=3$ for QCD).\par

In the case of the hybrid model, in between each splitting, the quark to gluon energy loss ratio is given by 
\begin{equation}\label{eq:energyratio_holo}
\frac{dE_{quark}}{dx}\bigg/ \frac{dE_{gluon}}{dx}=\left(\frac{C_F}{C_A}\right)^{\frac{1}{3}}=0.7631\cdots
\end{equation}
where the power law is introduced \emph{ad hoc}, based on holographic approaches to the energy loss puzzle \cite{Hybrid1,Hybrid2, Hybrid3}.\par

Our goal here is to isolate the vacuum-like emissions from medium-induced ones, knowing that both sources contribute to the total jet energy loss. An operational way of doing this is, as stated in section \ref{sec:1}, to introduce $R(\Delta p_T)_{med}$ where $\Delta p_T= \langle p_T^{Z}\rangle-\langle p_T^{jet}\rangle $. By averaging over all the events, we expect this variable to be strongly sensitive to the in-medium jet energy loss (in spite of jet by jet fluctuations), see eq. \eqref{eq:R}. In order to have a vacuum benchmark result we also compute $R(\Delta p_T)_{vac}$. As we will show, even for vacuum jets the Casimir scaling ($C_F/C_A$) is broken. This is expected and the deviation should come from the fact that (i) even to Leading Logarithmic accuracy (LL, i.e., when the jet is dominated by one perturbative gluon emissions), there are significant effects due to the fact that we are using a finite jet radius, R \cite{Dasgupta:2014} (ii) Monte Carlo results will include effects absent from LL calculations (for further discussion on this see \cite{Dasgupta:2007,Dasgupta:2014}). For completeness, we also show the results for $R(\Delta p_T)_{vac+med}$, that will refer to the full JEWEL or Hybrid simulation. All these results will allow us to explore how jet quenching modifies jets' colour structure beyond LL accuracy (which, as pointed out before, is naturally included in JEWEL and the Hybrid model).

\subsection{Parton level analysis}
\label{sec3.1}

\begin{figure}[h!]
    \centering
    \includegraphics[scale=0.65]{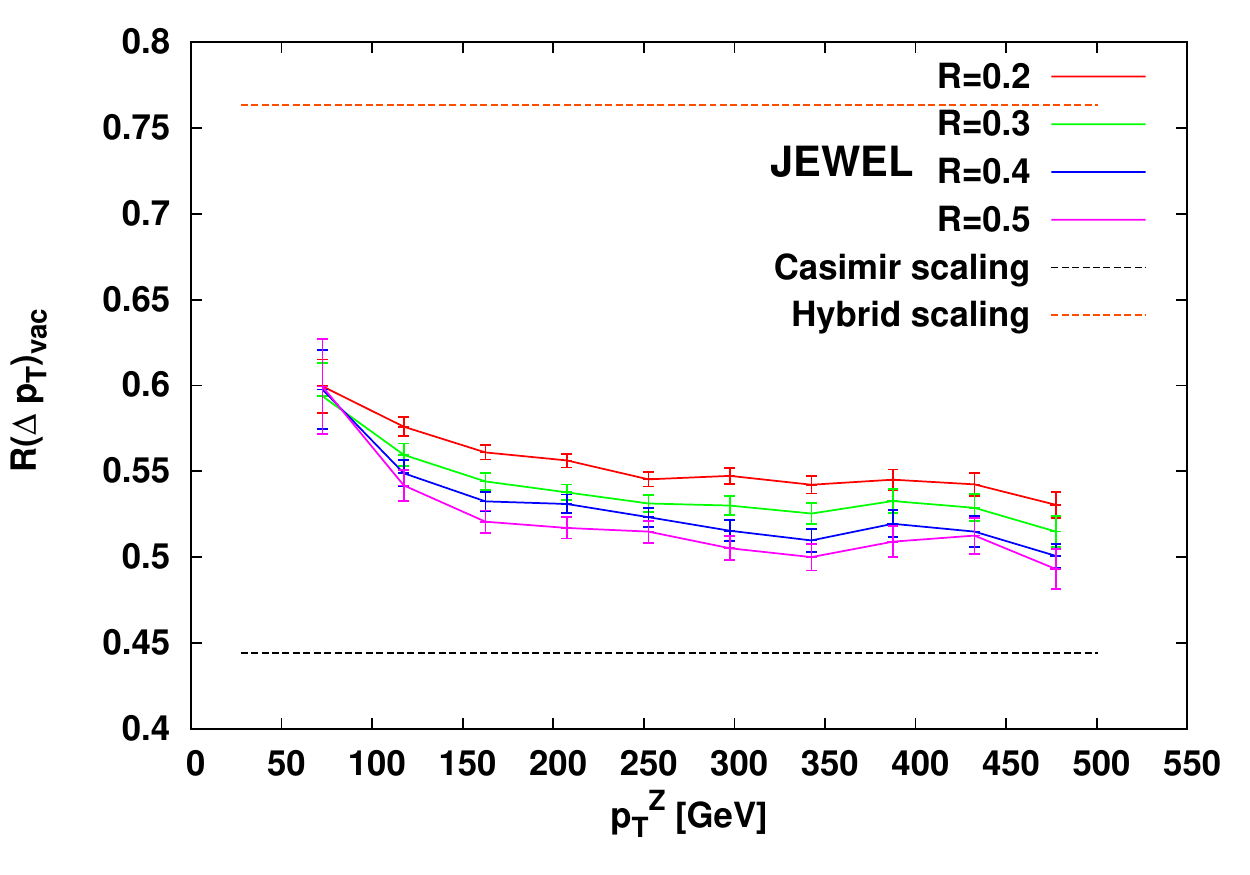}
    \caption{$R(\Delta p_T)_{vac}$ as a function of the initial parton $p_T$.}
    \label{fig:energyvac}
\end{figure}

We start by comparing $R(\Delta p_T)_{med}$ to $R(\Delta p_T)_{vac}$. In figure \ref{fig:energyvac} we plot the quark to gluon ratio in vacuum, $R(\Delta p_T)_{vac}$, as a function of the initial $p_T$ (identified as $p_T^Z$). In addition to the results for different jet radii $R$, we also provide two comparison guidelines: one exhibiting Casimir scaling and the other given by the Casimir's ratio to the power $1/3$ (hybrid scaling). We see that already for vacuum jets, Casimir scaling is not recovered. This observation is in line with the expectation that parton showers account for effects beyond the LL expectation. Nonetheless, for high enough $p_T$ jets, the scaling is approximately constant, between $0.50$ and $0.55$. 

\begin{figure}[h!]
    \centering
    \includegraphics[scale=0.65]{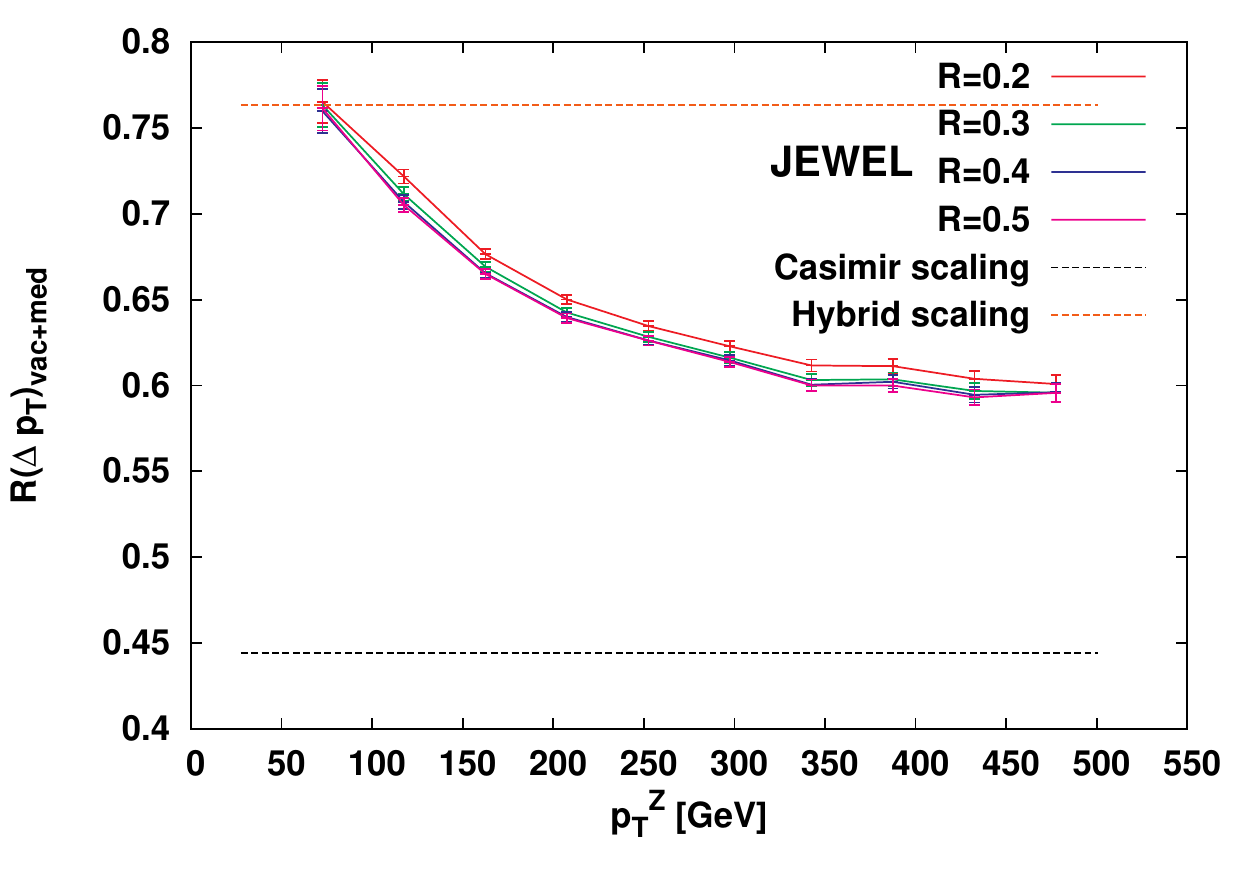}
    \caption{$R(\Delta p_T)_{vac+med}$ as a function of the initial parton $p_T$.}
    \label{fig:energymednew}
\end{figure}

\begin{figure}[h!]
    \centering
    \includegraphics[scale=0.65]{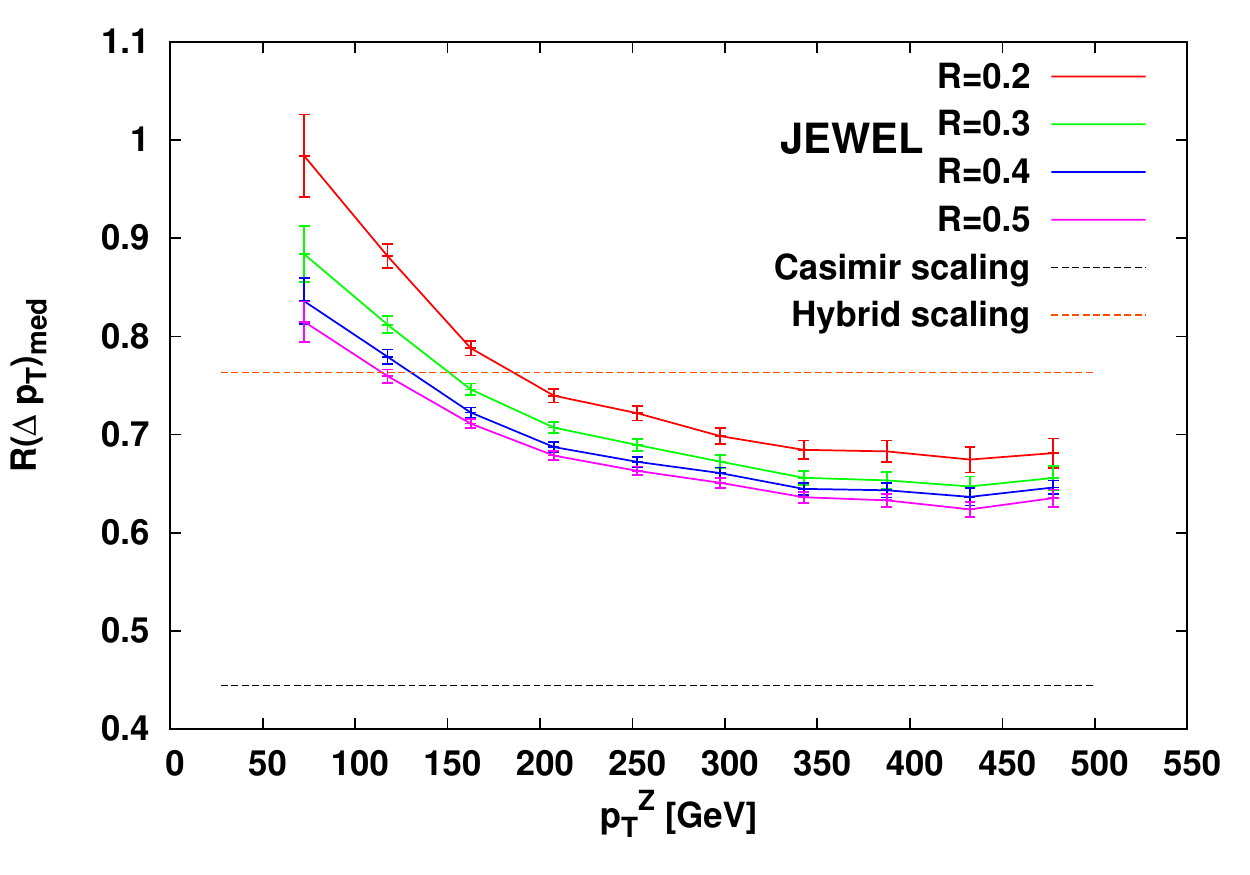}
    \caption{$R(\Delta p_T)_{med}$ as a function of the initial parton $p_T$.}
    \label{fig:energymed}
\end{figure}

In figures \ref{fig:energymednew} and \ref{fig:energymed}, $R(\Delta p_T)_{med}$ and $R(\Delta p_T)_{med+vac}$ distributions are shown, again as a function of $p_T^Z$. 

Before starting the discussion of our results it is important to point out some of the possible shortcomings of our approach. \par

Firstly, we point out a subtlety in our jet flavour selection. In this work, jet flavour is solely defined by the flavour of the initiating parton which is known since samples for quark and gluon initiated jets were generated separately. Such an operational definition can lead to a mistagging of the jet as it will not be sensitive to hard splittings at early times in which the largest momentum fraction is carried by a parton with a different flavor from the parent one. JEWEL relies on leading order hard matrix elements for the generation of events where a jet recoils against a $Z-$boson. As such, within JEWEL, there is no ambiguity regarding the flavour of the (single) initial parton.  Additionally, the Altarelli-Parisi splitting functions (at LO) are highly suppressed for flavour changing configurations (quark pair creation) and gluon hard emissions off quark. These arguments are therefore sufficient to guarantee that our approximation still provides a robust operational definition. \par

Another important aspect in the discussion is related to the definition of jet energy loss in the presence of (colour) decoherence \cite{coherence_colour}. As stated in section \ref{sec:1}, the identification of the jet energy loss with the energy loss of the incoming parton is only valid, qualitatively, when the jet is completely coherent. In the explored kinematic range ($p_T^Z\geq 200 \ GeV$), jets are expected to behave within the two limits (fully coherent or decoherent evolution of the respective parton shower). JEWEL only has a minimal implementation of colour decoherence while the Hybrid model is based on the vacuum parton shower evolution as provided by Pythia. Nonetheless, such effects should be important to determine the effective number of colour charges inside of the jet. Consequently, the concept of jet energy loss and jet colour charge are entangled with the presence of decoherence. A further study of this effect alone also needs an operational way of disentangling decoherent and coherent partons, which, so far, was not possible. We therefore do not take this effect into account when discussing the results throughout the manuscript. With these caveats, we observe an interesting result in figures \ref{fig:energymednew} and \ref{fig:energymed}: for high $p_T$ bins, Casimir scaling is broken for the medium induced emissions with the same jet radius dependence as in vacuum, but scaled by a multiplicative constant. Remarkably, the pure medium-induced Casimir scaling (fig.\ref{fig:energymednew}) is larger than the vacuum ($\approx 1.25-1.30$ times), which means that quark-initiated jets and gluon-initiated jets radiate quite similarly in the presence of a medium. This is important for phenomenological applications since it shows that the correct quark to gluon-initated jets scaling for medium induced radiation is larger than the original expectation (Casimir scaling). \par 

From figure \ref{fig:energymed}, we can also check that for the (pure) medium-induced emissions the ratio obtained is significantly larger than the one obtained in figure \ref{fig:energyvac}.\par 

As for the low $p_T$ bins, we observe that the evolution is rather quick to the asymptotic value (value reached in the high $p_T$ bins), although its absolute value depends on the setup. These bins are more sensitive to jet by jet fluctuations (since there are less partons inside each jet). Despite this different behaviour in transverse momentum, the evolution is decreasing with $p_T^Z$, thus ensuring that energetic jets will all show the same scaling.\par

\begin{figure}[h!]
    \centering
    \includegraphics[scale=0.65]{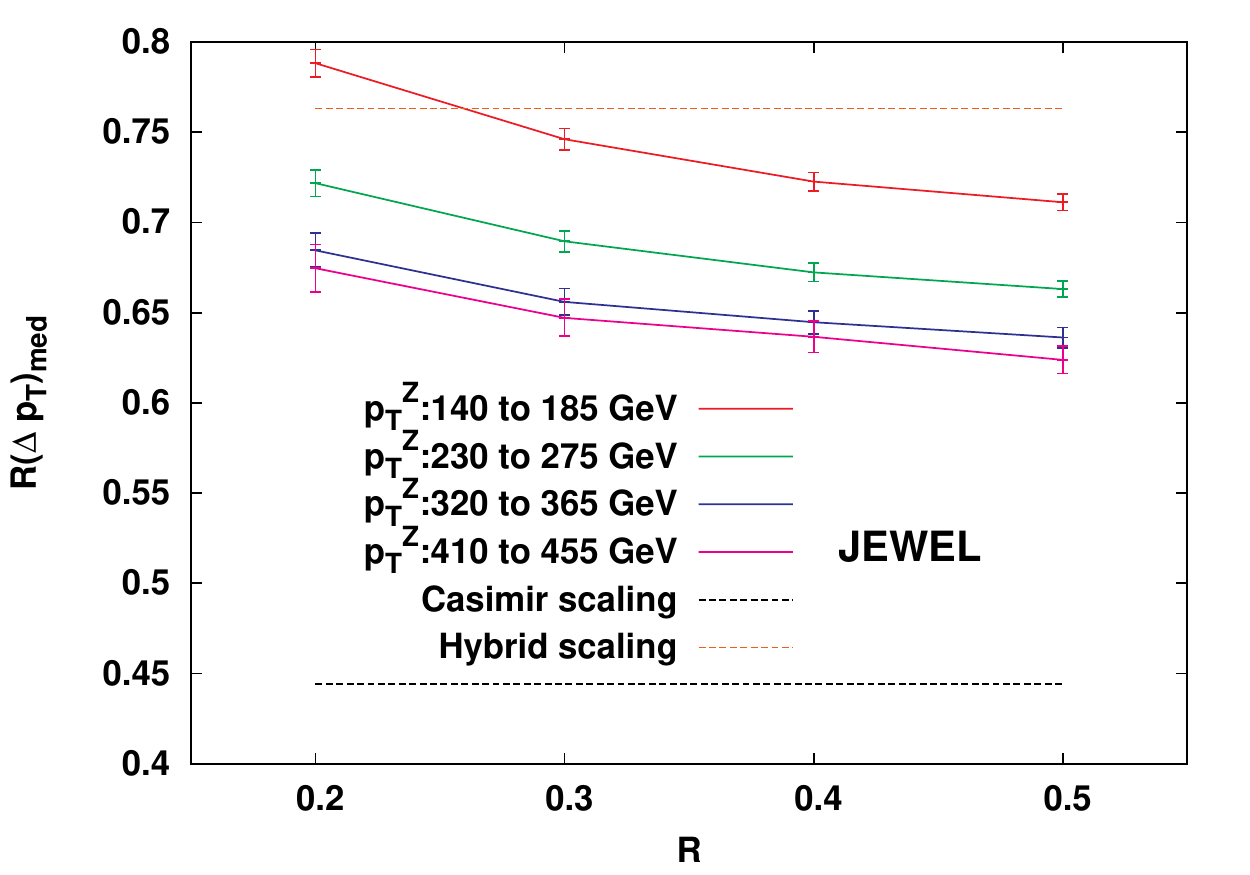}
    \caption{$R(\Delta p_T)_{med}$ as a function of the reconstructed jet radius.}
    \label{fig:energymedR}
\end{figure}

Interestingly, when medium-induced and vacuum-like emissions are put together in the parton shower, there is still a constant scaling behaviour for energetic jets, but the jet radius dependence is now very small. In fact, the evolution with the jet radius $R$, figure \ref{fig:energymedR}, shows that for high $p_T$ jets the medium induced scaling is roughly insensitive to the jet radius (the dependence decreases faster than linearly). From figure \ref{fig:energymedR}, it is also possible to see that quark-initiated jets and gluon-initiated jets are more alike when the jet radius decreases. This might seem counter-intuitive, as one might expect that for small jet radius, one would recover the single particle behaviour. Along this line, the ratio would decrease, approaching Casimir scaling. However, small jet radius have a non-trivial dependence on $R$ (see  \cite{Dasgupta:2014}, which exhibits the same $R$ ordering as we obtained).

\subsection{Hadronization and Initial State Radiation analysis}
\label{sec3.2}

So far, the analysis was made for an ideal environment, at parton level. We now proceed to analyse the samples with hadronization and initial state radiation, in order to see how the scaling above gets modified. In particular, we expect ISR to affect the vacuum Casimir scaling since it will both populate the jets with extra radiation and introduce jets in the sample that have not been produced back-to-back with a $Z-$boson. Following the study carried out in \cite{Milhano:2015mng} we expect both this contribution and the effect of hadronization to be small.
Also, any remaining effects of hadronization should cancel out between the vacuum and medium contributions in $R(\Delta p_T)_{med}$.\par  

\begin{figure}[h!]
    \centering
    \includegraphics[scale=0.65]{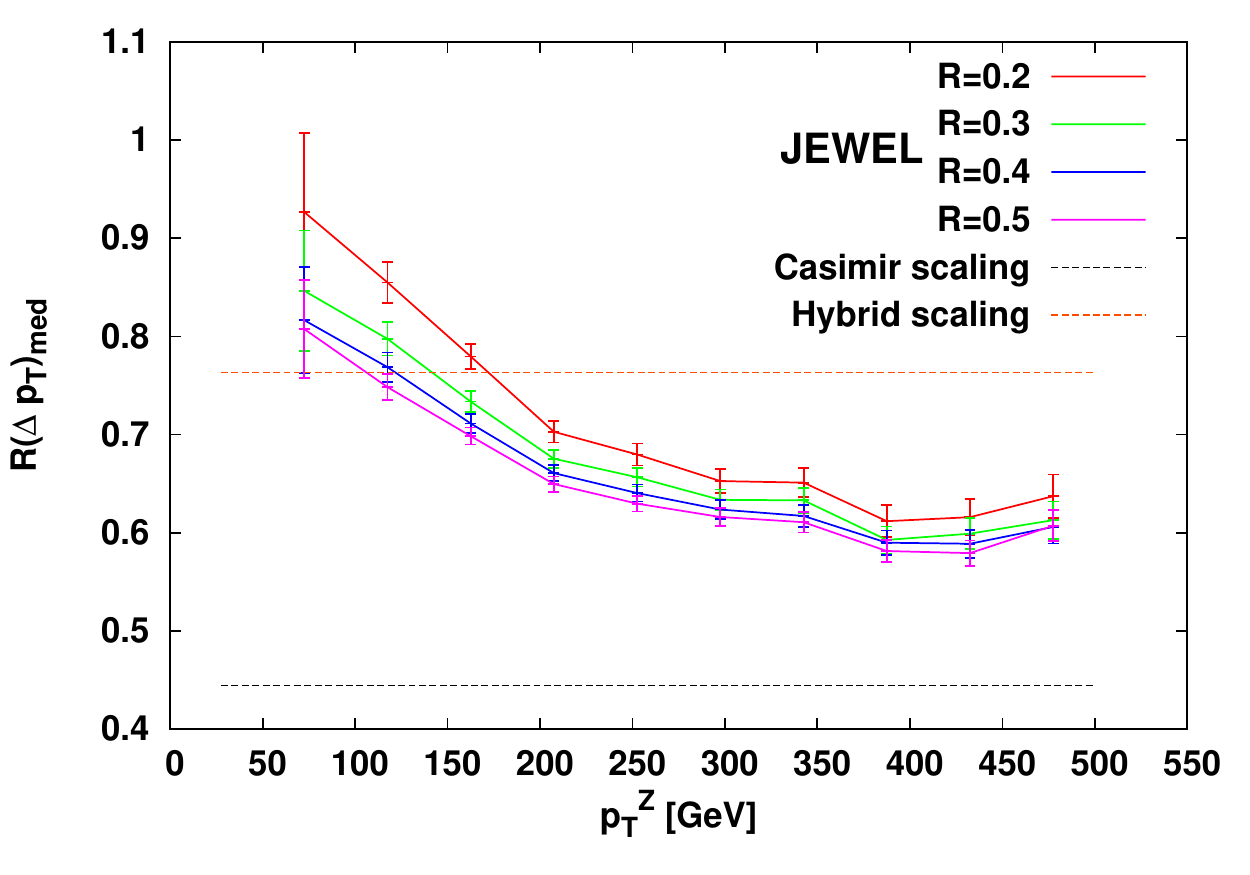}
    \caption{$R(\Delta p_T)_{med}$ as a function of the initial parton $p_T$ for the samples with hadronization.}
    \label{fig:Henergymed}
\end{figure}

\begin{figure}[h!]
    \centering
    \includegraphics[scale=0.65]{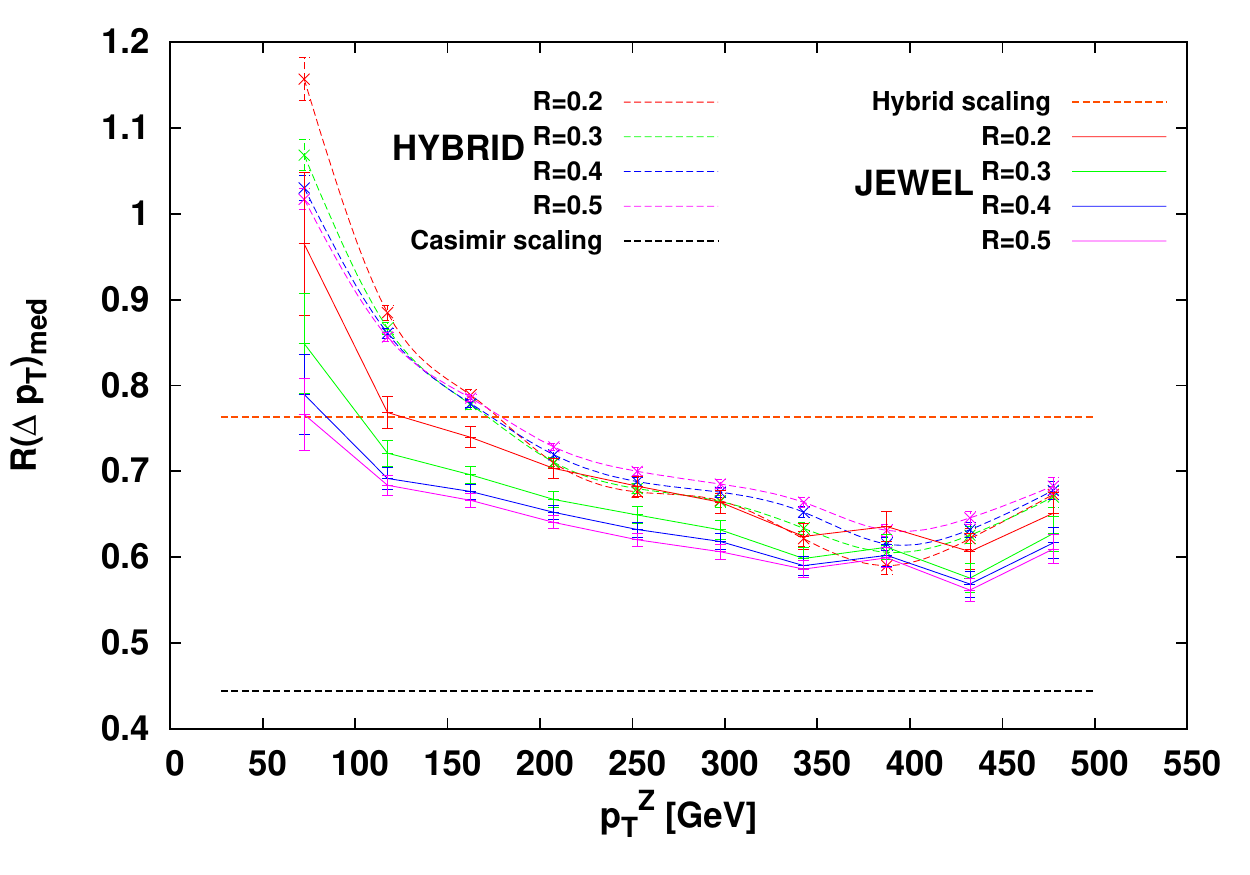}
    \caption{$R(\Delta p_T)_{med}$ as a function of the initial parton $p_T$ for the samples with ISR. The Hybrid model samples are shown with spline lines are correspond to a $0-10 \ \%$ centrality events.}
    \label{fig:ISRenergymed}
\end{figure}

Figure \ref{fig:Henergymed} summarises the hadronization results while figure \ref{fig:ISRenergymed} summarises the results when only ISR is included. In this case, we also show, for direct comparison, the results obtained from the Hybrid approach (recall that ISR and hadronization are activated separately in each sample of either model).

Despite the slightly fluctuating results, we observe that, qualitatively, the behaviour is again a simple rescaling of the vacuum Casimir law. We note however that the stable high $p_T$ values tend to be lower than the partonic samples. 
We also want to point out that the initial state radiation result shows smaller uncertainties and fluctuations. This might be due to the fact that hadronization is directly sensitive to the medium effects while initial state radiation is not. \par 

From figure \ref{fig:ISRenergymed}, it is clear that both generators show stable ratios for high $p_T^Z$ at fairly similar values. This observation is rather remarkable since the colour structure of the showers in the two models is very different. In JEWEL the colour structure of the shower is modified by interaction with medium scattering centres. In particular, by each gluon exchanged with the coloured medium, the incoming parton has its colour field rotated. The Hybrid model preserves the colour structure of the PYTHIA generated shower. The only modification of colour structure is induced by the partons that are stopped in the medium due to energy loss effects and thus will not contribute to the reconstructed jet. Taking into account the holographic scaling from \eqref{eq:energyratio_holo}, we would naively expect quark- and gluon-initiated jets to be more alike in the Strong/Weak Hybrid model than in  JEWEL. We see in figure \ref{fig:ISRenergymed} that this is indeed the case, but the difference is not significant. One of the contributing factors might be the presence of additional radiation sources (particles) that naturally occur during JEWEL parton shower evolution, with respect to Hybrid (where energy losses are applied to a fully developed vacuum-like shower). From our studies in section \ref{sec3.1}, the convolution of pQCD medium-induced energy loss, together with some decoherence effects during the parton shower evolution, induce similar energy losses for quark- and gluon-initiated jets. Overall, the JEWEL Casimir ratio increases and approaches the one obtained from the Hybrid model.\par

At sufficiently high $p_T$ the $R$ ordering of the Hybrid model curves is the opposite to what JEWEL predicts. For both JEWEL and the Hybrid model, at low $p_T$, the behaviour is dominated by gluon-like (which radiate more), thus we observe a larger value of $R(\Delta p_T)_{med}$. At larger $p_T$, quark-like jets dominate and thus the value of $R(\Delta p_T)_{med}$ decreases. For a fixed $p_T$ bin (sufficiently high), increasing $R$ leads to recovery of radiation for both quark- and gluon-like jets. However, this is more relevant for gluon-like jets, and therefore the scaling decreases with $R$ in JEWEL. However, in the Hybrid model, what happens is that in fact, a part of the jet population (for a certain $p_T$ bin) comes from jets which lost a lot of energy, and thus migrated from higher $p_T$ bins. Therefore, as one changes the radius, there is a competition between a population of jets which did not lose much energy $vs$ a population of higher energy jets which radiated a lot and ended in a lower $p_T$ bin. From the behaviour seen as $R$ varies, we see that this second group significantly modifies the dependence on $R$, when comparing to JEWEL.\par

\subsection{Centrality results}
\label{sec3.3}

For completeness, we analyse the evolution of the Casimir ratio as a function of centrality. For that, we include the two additional samples corresponding to $[10-20]\%$ and $[30- 50]\%$ centralities.\par 

\begin{figure}[h!]
    \centering
    \includegraphics[scale=0.65]{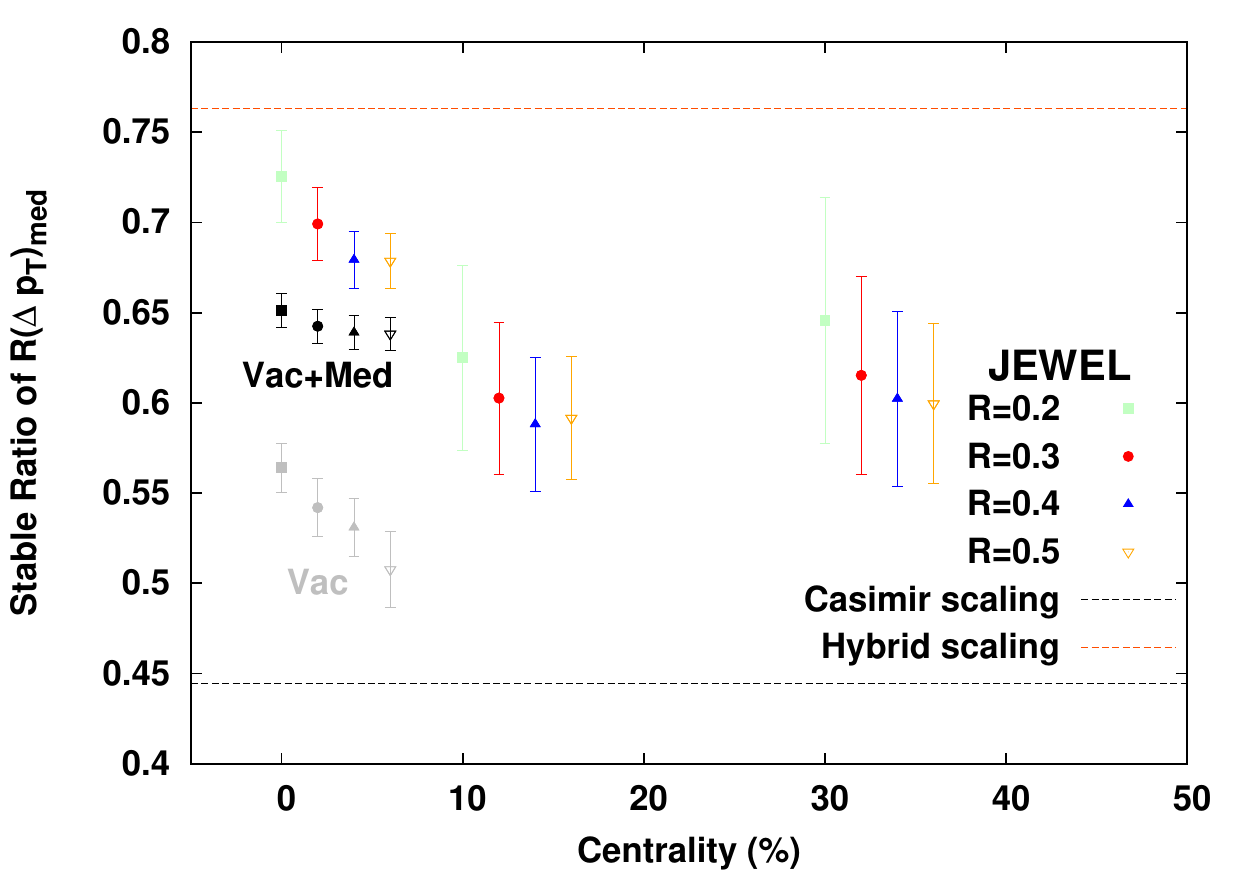}
    \caption{The asymptotic scaling extracted via fitting $R(\Delta p_T)_{med}$ in $p_T^Z$. We plot the data points in sets of 4 ($0-10 \ \%$, $10-20 \ \%$ and $30-50\ \%$). For each centrality bin, the points of different jet radius have an offset in centrality for a better visualisation of the results. Black points labeled \textit{medium} correspond to the extraction from $R(\Delta p_T)_{vac+med}$ and grey points labeled \textit{vacuum} were extracted from $R(\Delta p_T)_{vac}$.}
    \label{fig:stable_ratios}
\end{figure}

For a better comparison among all results, we fit the last five bins of $R(\Delta p_T)_{med}$ in $p_T^Z$ to a linear model
\begin{equation}\label{eq:linear_fit}
R(\Delta p_T)_{med}[p_T^Z]=a+b \ p_T^Z    
\end{equation}
We find that for all the samples the best fit shows a very small slope $b$ (at least three orders of magnitude smaller than the constant $a$ fitted). Therefore, we assume a constant value for highly energetic jets we used that extracted value as the \emph{Stable Ratio}. The results are shown in figure \ref{fig:stable_ratios}, where reach ratio was slightly displaced within the centrality bin to improve readability. The colourful results refer to $R(\Delta p_T)_{vac+med}$, and we leave, for reference, the $R(\Delta p_T)_{med}$ (in black) and $R(\Delta p_T)_{med}$ (in gray).

For the new samples we find a larger uncertainty in the extraction of this value and looking directly to the scaling in $p_T^Z$ we see that the results are not as stable as the $[0-10] \ \%$ sample. For the largest centrality, this is reflected in very large error bars. \par 

As expected, as the centrality increases a decrease in the calculated quark to gluon ratio but always above the observed value for the vacuum sample. This trend is not seen in the largest centrality bin. The values for this centrality class are higher than expected. This might be due first to the lower statistics available and secondly, by direct observation of $R(\Delta p_T)[p_T^Z]$, we see that as the centrality increases, this linear fit becomes less accurate and the sample takes more bins to stabilise. Therefore, to properly extract the correct scaling for this class one would need to include higher $p_T$ bins. In addition, it is known that the JEWEL centrality evolution is not correctly captured (for instance for $R_{AA}$).

\section{Conclusion}\label{sec:4}

In this manuscript we have studied the role that jet quenching plays in the modification of the net colour charge of QCD jets. We have performed a broad study, exploring two separate Monte Carlo generators, to account for pQCD and Holographic estimates of the Casimir quark-to-gluon ratio. \par

In section \ref{sec:2} we presented the major results of this paper, where we relied on the variable $R(\Delta p_T)_{med}$. As argued before, it should be essentially sensitive only to the medium-induced energy loss (by $p_T^Z$ bin). Our results confirm the breaking of the traditional vacuum Casimir scaling by an overall multiplicative constant. Moreover, such a scaling violation leads to quark- and gluon-initiated jets being more similar in medium than in vacuum.\par 

\begin{figure}[h!]
    \centering
    \includegraphics[scale=0.65]{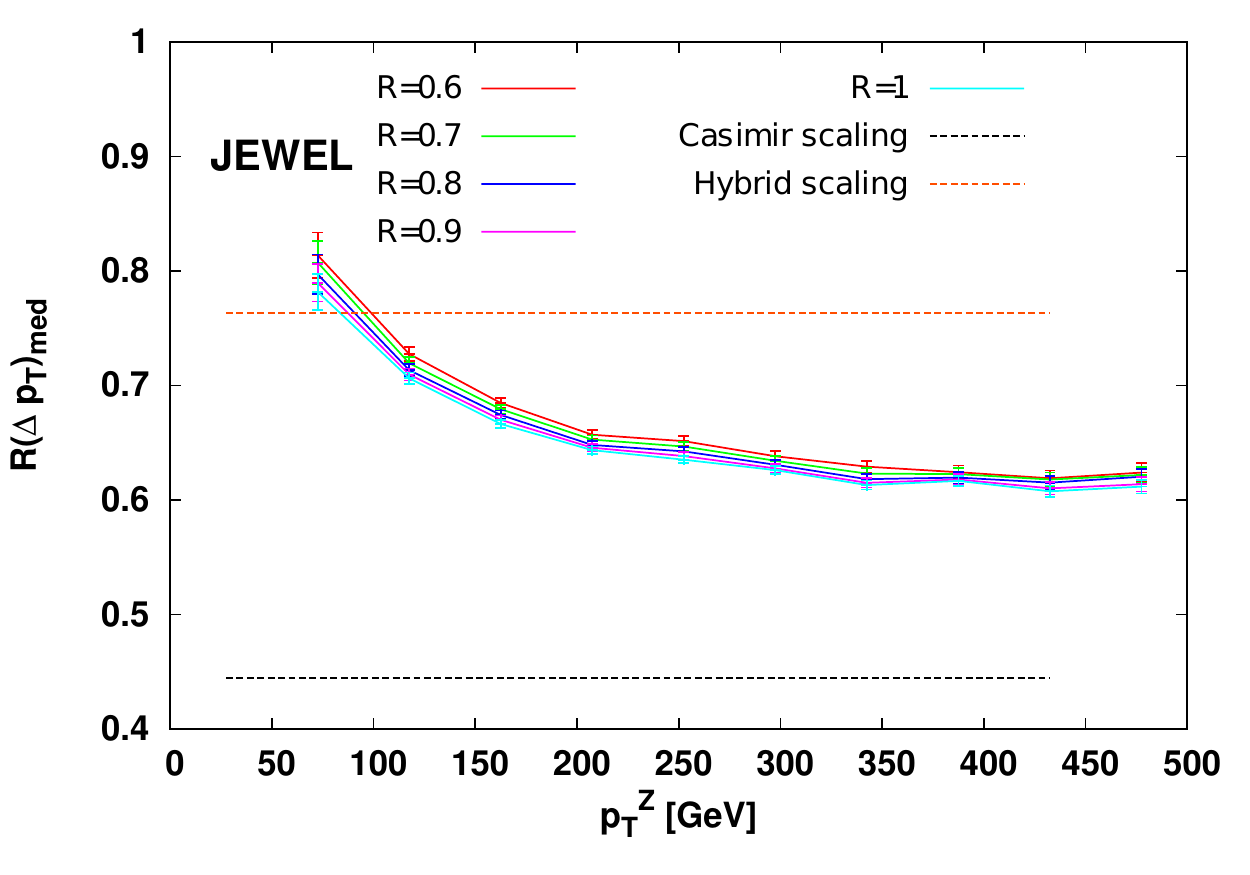}
    \caption{Scaling for large jet parameter values of the scaling just for $R(\Delta p_T)_{med}$.}
    \label{fig:full_largeR}
\end{figure}

\begin{figure}[h!]
    \centering
    \includegraphics[scale=0.65]{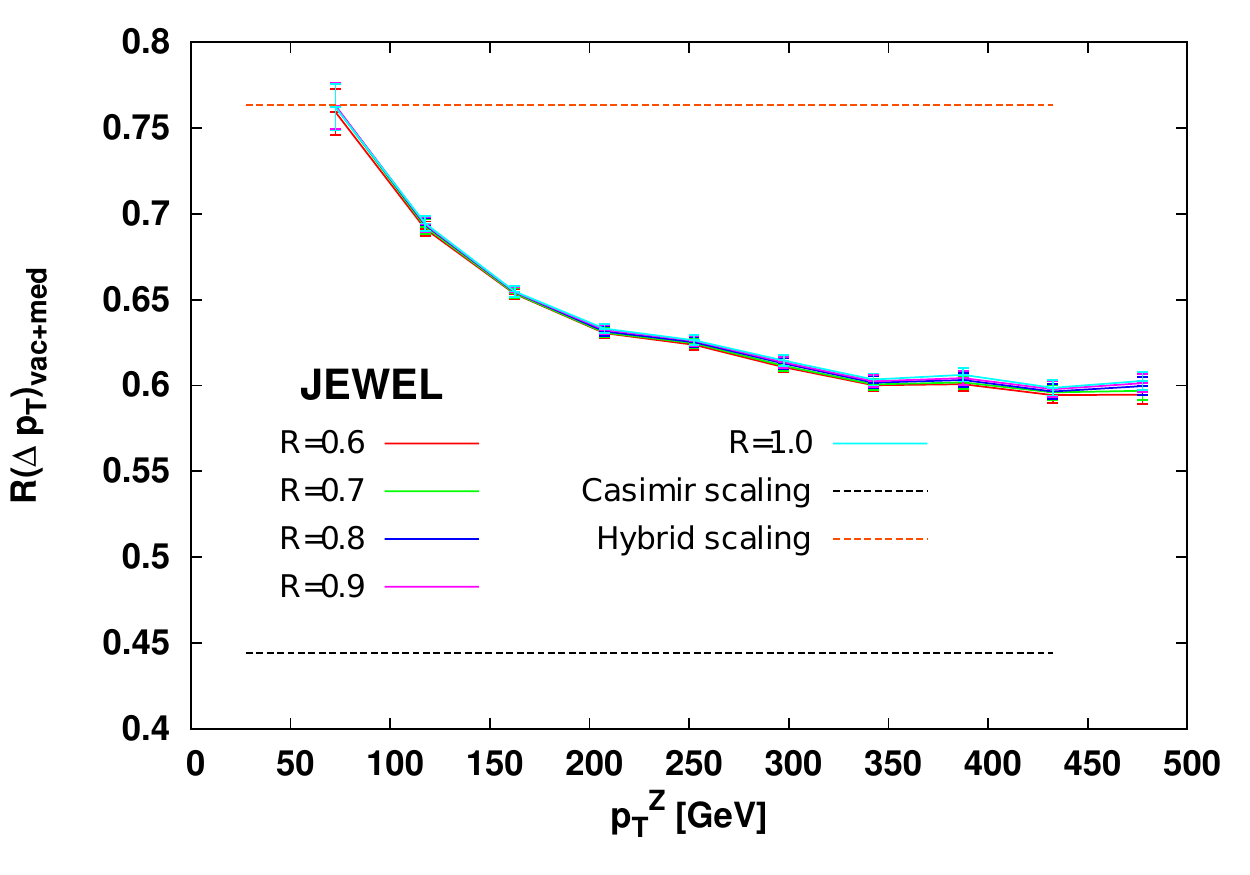}
    \caption{Scaling for large jet parameter values of the scaling just for $R(\Delta p_T)_{vac+med}$.}
    \label{fig:medium_largeR}
\end{figure}

The scaling obtained tends to converge at high $p_T$ and becomes insensitive to the jet radius $R$. In addition, our results show that (soft medium induced) radiation is transmitted to gluons at very large angles (bigger than $R=0.5$). For completeness, we extended our analysis to radius between $R=0.6$ and $R=1$ (figures \ref{fig:full_largeR} and \ref{fig:medium_largeR}). We see that the results follow the trend seen at smaller $R$ and confirm the universality of the property in study.\par 

Our results also reinforce the picture that jet and parton energy loss are different processes and one cannot be obtained from the other by a simple rescaling law. This observation is at the heart of developing a consistent and well posed framework for jet quenching and thus we reiterate the message that for phenomenological studies of jets a non trivial implementation of the energy loss mechanism is required. \par 

In the literature, as far as we are aware, no attempt of studying Casimir scaling for in-medium jets had ever been attempted. Nonetheless, this quantity is extensively used when, for example, doing fits to data. Two such direct applications appear in \cite{Spousta,SpoustaCole}. In the first of these two papers a global fit allows one to obtain the parameter $c_F=1.78\pm0.12$. This should correspond to the inverse of our scaling, thus corresponding to $1/c_F=0.56\pm 0.04$. This value is slightly smaller than the one shown in figure \ref{fig:stable_ratios}. The value extracted in the second paper is of $O(4)$, thus in complete in disagreement with our findings. Such a disparity might come from the complex nature of the fitting procedure in \cite{SpoustaCole}. It should be mentioned that the value quoted refers to the $[0-1] \ \%$ centrality class. Nonetheless, an interesting characteristic of these fits is that they both find that the data is well described if Casimir scaling is broken by a multiplicative factor. This suggests that a constant law for the quark to gluon ratio is pretty robust. More recently, a theoretical effort \cite{YacineSoeren} also found that just for the medium induced cascade, extended showers should depart from Casimir scaling. \par
In the future and for applications involving data, we expect the results of this paper to show the importance of using the correct Casimir scaling for in-medium jets. Nonetheless, we would like to point out that any application would require a more controlled approach, with cross checks using other observables. 

Finally, for a full picture of Casimir scaling in the medium, a better and more complete understanding of jet inner structure is needed, since the scaling will depend on the finer details of each jet and not just the jet radius and the $p_T$ of the initiating parton.

\begin{acknowledgements}
We are grateful to Alba Soto Ontoso, Carlos Salgado and N\'estor Armesto for useful discussions. We thank Daniel Pablos for providing the Hybrid model samples which made our study possible.
 The project that gave rise to these results received the support of a fellowship from ``la Caixa" Foundation (ID 100010434). The fellowship code is LCF/BQ/ DI18/11660057. This project has received funding from the European Union's Horizon 2020 research and innovation programme under the Marie Sklodowska-Curie grant agreement No. 713673.
  LA acknowledges the financial support by OE - Portugal, FCT, I.P., under DL/57/2016/CP1345/ CT0004.
JB is supported by Ministerio de Ciencia e Innovacion of Spain under project FPA2017-83814-P; Unidad de Excelencia Maria de Maetzu under project MDM-2016-0692; European research Council project ERC-2018-ADG-835105 YoctoLHC; and Xunta de Galicia (Conselleria de Educacion) and FEDER. JB is supported in part at BNL under DOE grant de-sc0012704. JB also acknowledges the support from the Fulbright Comission. LA and JGM are supported by Funda\c{c}\~{a}o
para a Ci\^encia e a Tecnologia (Portugal) under project CERN/FIS-PAR/0022/2017. JGM gratefully acknowledges the hospitality of the CERN theory group.
\end{acknowledgements}


\bibliographystyle{spphys}       
\bibliography{Lib.bib}   
a

\end{document}